\documentclass[twocolumn,pre,amssymb,amsmath,natbib,showpacs,floatfix,superscriptaddress]{revtex4}%

\usepackage[dvips]{graphicx}
\usepackage{color}

\newcommand{\Eref}[1]{{Eq.~(\ref{#1})}}
\newcommand{\Fref}[1]{{Fig.~\ref{#1}}}

\newcommand{\rmi}{{\rm i}}
\newcommand{\rmd}{{\rm d}}

\begin{document}

\title{Fidelity decay in interacting two-level boson systems: Freezing and revivals}

\author{Luis Benet}%
\thanks{On leave at Dept. de Matem\`atica Aplicada i An\`alisi, Universitat de Barcelona, Spain.}
\affiliation{Instituto de Ciencias F\'{\i}sicas, Universidad Nacional Aut\'onoma de M\'exico (UNAM), Cuernavaca, M\'exico}%
\affiliation{Centro Internacional de Ciencias, Cuernavaca, M\'exico} 
\author{Sa\'ul Hern\'andez-Quiroz}
\affiliation{Instituto de Ciencias F\'{\i}sicas, Universidad Nacional Aut\'onoma de M\'exico (UNAM), Cuernavaca, M\'exico}%
\affiliation{Facultad de Ciencias, Universidad Aut\'onoma del Estado de Morelos, 
  Cuernavaca, M\'exico}%
\author{Thomas H. Seligman}
\affiliation{Instituto de Ciencias F\'{\i}sicas, Universidad Nacional Aut\'onoma de M\'exico (UNAM), Cuernavaca, M\'exico}%
\affiliation{Centro Internacional de Ciencias, Cuernavaca, M\'exico} 
  
\date{\today}

\begin{abstract}
We study the fidelity decay in the $k$-body embedded ensembles of random 
matrices for bosons distributed in two single-particle states, considering the 
reference or unperturbed Hamiltonian as the one-body terms and the diagonal 
part of the $k$-body embedded ensemble of random matrices, and the perturbation 
as the residual off-diagonal part of the interaction. We calculate the ensemble-averaged 
fidelity with respect to an initial random state within linear response theory to second 
order on the perturbation strength, and demonstrate that it displays the freeze of the 
fidelity. During the freeze, the average fidelity exhibits periodic revivals at integer values 
of the Heisenberg time $t_H$. By selecting specific $k$-body terms of the residual 
interaction, we find that the periodicity of the revivals during the freeze of fidelity is an 
integer fraction of $t_H$, thus relating the period of the revivals with the range of the 
interaction $k$ of the perturbing terms. Numerical calculations confirm the analytical 
results.
\end{abstract}


\maketitle

\section{Introduction}

Fidelity, or Loschmidt echo, was introduced to study the effect of perturbations on the 
dynamics of quantum systems, mainly in the context of quantum information, but also 
in the study of the stability of quantum systems; for a review see~\cite{Gorin2006}. Prosen 
and \v{Z}nidari\v{c} noted that a freeze of fidelity occurs, i.e. fidelity will remain stable 
for very long times in the scale of the Heisenberg time, if the diagonal part of the 
perturbation matrix vanishes~\cite{ProsenZnid2003}. This lead to an additional 
interesting view on fidelity decay. In any given system, we can view any part of the 
Hamiltonian as the {\it unperturbed system} and the rest as {\it perturbation}. In 
particular in a many-body system the mean-field theory can be considered to be the 
unperturbed system, and the residual interaction as the perturbation. Fidelity thus 
serves as a measure of the quality of such mean-field approach. This was actually 
considered in Ref.~\cite{Pizorn2007} in the context of a random matrix model using a 
two-body random ensemble, i.e. an ensemble that takes into account the two-body 
character of the interactions as well as the fermionic character of the particles. The 
result was rather surprising in the sense, that the freeze does not occur for the 
ensemble-average of fidelity, but it does occur for the median fidelity, or equivalently 
for the average of the logarithm of fidelity, also known as 
distortion~\cite{Lobkis2003,GSW2006}. In the bosonic case we can ask a similar 
question, and we have the pleasant situation that such systems are Liouville 
integrable in the semiclassical limit if the bosons are restricted to two 
levels~\cite{BJL2003}. This special case is attractive since it is experimentally 
accessible~\cite{Anderson1998,Albiez2005,Morsch2006,Gati2007}. 

In the present paper we shall analyze this case in the framework of the embedded 
random matrix ensembles for bosons, i.e., we will allow in principle also three-body 
and higher-order interactions. Such ensembles have a long history for fermion 
systems~\cite{fre70,boh71,mon75} and a somewhat shorter one for 
bosons~\cite{kot80,man84,Asaga2001}. More recently such ensembles have even 
been defined for distinguishable particles~\cite{Pizorn2008}. For recent reviews 
see~\cite{Kota2001,BW2003}.
For the bosonic case the freeze of fidelity is readily seen~\cite{BHQS2010}, 
if we use the simplistic version of a mean-field theory that in addition includes the 
diagonal part of the residual interaction (in the representation in which the one-body 
component of the Hamiltonian is diagonal). Besides the freeze of the fidelity, we also 
uncover unexpected revivals at fractions of the Heisenberg time, which reflect the 
many-body residual interactions. 

The paper is organized as follows: In Section~\ref{sec:Definitions} 
we define the $k$-body two-level bosonic ensemble of random matrices and fidelity, 
for which the reference and perturbed Hamiltonians are fully specified. In 
Section~\ref{sec:EAverag} we calculate within the linear response theory the ensemble 
average of the fidelity and find that it is a Fourier cosine-series whose basic periodicity 
is the Heisenberg time. In Section~\ref{sec:PFrevivals} we carefully select the 
perturbing off-diagonal $k$-body terms and obtain fractional periods of the revivals 
during freeze, in units of the Heisenberg time. In Section~\ref{sec:Concl} we summarize 
our work and outline the conclusions.

\section{Definitions}
\label{sec:Definitions}

\subsection{The $k$-body two-level bosonic ensemble of random matrices}
\label{sec:BEE}

To define the $k$-body Bosonic Embedded Ensemble of Random Matrices for 
bosons~\cite{Asaga2001,BW2003}, we consider $n$ spin-less
bosons distributed over $l$ single-particle states. These bosons are associated
with the creation and annihilation bosonic operators $\hat a^\dagger_j$ and 
$\hat a_j$, respectively, with $j=1,\dots, l$. 
From here on, we shall focus on the two-level case $l=2$, which is the simplest 
one, has some remarkable properties~\cite{BJL2003,BLS2003,HQ-B2010}, 
and is also interesting from an experimental point of view for the two-component 
Bose-Einstein condensates~\cite{Morsch2006,Gati2007}. 

We denote the normalized two-level $n$-boson states as 
$| \mu \rangle \equiv | \mu, n-\mu \rangle = ({\cal N}_{\mu,n})^{-1} 
(\hat a_1^\dagger)^\mu (\hat a_2^\dagger)^{n-\mu}|\vec{0}\rangle$, 
where ${\cal N}_{\mu,n}=[\mu! (n-\mu)!]^{1/2}$ is a normalization constant and 
$| \vec{0}\rangle$ is the vacuum state. This is the 
occupation-number basis spanned by $\mu=0,\dots n$; the Hilbert--space dimension 
is thus $N=n+1$. These states are coupled through a random $k$-body interaction 
$\hat H_k({\beta})$, which is written as~\cite{BLS2003,HQ-B2010}
\begin{equation}
  \label{eq:Hk}
  {\hat H_k (\beta)} = \sum_{r,s=0}^k \, v_{r,s}^{(\beta)} \,
  \frac{ (\hat a_1^\dagger )^{r} ( \hat a_2^\dagger )^{k-r} 
    (\hat a_1 )^{s} ( \hat a_2 )^{k-s} }{{\cal N}_{r,k}\, {\cal N}_{s,k} } \, .
\end{equation}
Here, $k$ denotes the rank of the interaction, $1\le k \le n$, and $v_{r,s}^{(\beta)}$
are the $k$-body matrix elements, which are independent Gaussian-distributed 
random numbers with zero mean and constant (fixed) variance $v_0^2=1$. As in 
the case of the canonical random matrix ensembles~\cite{GMGW1998}, Dyson's 
parameter $\beta$ distinguishes the cases according to time-reversal invariance: 
$\beta=1$ is the time-reversal symmetric case, and $\beta=2$ is the case where 
this symmetry is broken. Hence, the $k$-body interaction matrix $v^{(\beta)}$ is a 
member of the Gaussian orthogonal ensemble (GOE) for $\beta=1$, or Gaussian 
unitary ensemble (GUE) for $\beta=2$. Notice that $\hat H_k (\beta)$ commutes 
with the number operator 
${\hat n}=\hat a_1^\dagger \hat a_1 + \hat a_2^\dagger \hat a_2$, i.e., the interaction 
conserves the total boson number $n$. 

As mentioned above, this ensemble for $l=2$ presents some noteworthy properties: 
It exhibits non-ergodic level statistics in the dense limit~\cite{Asaga2001}, 
i.e., spectral or ensemble unfolding do not yield the same results when $k$ is fixed 
and $n\to\infty$. In addition to this, for $\beta=1$ the ensemble displays 
a large and robust quasi-degenerate portion of the spectrum for a wide range of $k$, 
the Shnirelman doublets~\cite{Shnirelman1975}, while for $\beta=2$ only seldom 
accidental quasi-degeneracies are observed~\cite{HQ-B2010}. These results 
are consistent with the fact that each member of the ensemble is Liouville integrable 
in the semiclassical limit~\cite{BJL2003}.

For later purposes, we write the $k$-body Hamiltonian as $\hat H_k (\beta) = 
\hat H_{0_k}({\beta})+\hat V_k (\beta)$, which are defined by
\begin{equation}
\label{eq:split}
\langle \mu | \hat H_k (\beta) | \nu \rangle  = 
\langle \mu | \hat H_{0_k}({\beta}) | \nu \rangle \delta_{\mu,\nu}
+\langle \mu | \hat V_{k} (\beta) | \nu \rangle (1-\delta_{\mu,\nu}).
\end{equation}
Therefore, in the occupation-number basis, $\hat H_{0_k}({\beta})$ is the diagonal 
part of $\hat H_k (\beta)$, and $\hat V_k (\beta)$ contains only the 
off-diagonal contributions.

\subsection{Reference and residual interactions}

Fidelity is a measure of stability for small changes of a reference 
Hamiltonian~\cite{Gorin2006}. Therefore, we must begin by defining the 
reference or unperturbed Hamiltonian $\hat{\cal H}_0$, and the perturbed one,
which we write as $\hat{\cal H}_\lambda = \hat{\cal H}_0 + \lambda\hat{\cal V}$,
where $\lambda$ is the perturbation strength.
We shall be interested in the case where the total interaction consists of
a diagonal one-body interaction coupled to the $k$-body two-level bosonic 
embedded ensemble. In particular we shall study the cases $k=2$ or $k=3$ which 
are physically the most relevant.

To this end, we shall focus on a specific choice of $\hat{\cal H}_0$ and $\hat{\cal V}$, 
where we consider that $\hat{\cal V}$ has zero diagonal elements in the eigenbasis 
of the unperturbed Hamiltonian $\hat{\cal H}_0$; in this case, we shall refer to 
$\hat{\cal V}$ as the residual interaction. Notice that this case mimics the 
typical set-up in mean-field calculations, though it is also encountered in other 
cases of physical interest, e.g., when the perturbation is a time-reversal symmetry 
breaking interaction~\cite{GKPSSZ2006}. We are interested in this type of reference 
and the residual interactions since they fulfill the conditions to observe the fidelity 
freeze~\cite{ProsenZnid2003}, which implies longer stability times. 
Therefore, we include the diagonal part of the $k$-body interaction in the definition 
of the reference Hamiltonian, which we write as
\begin{equation}
\label{eq:H0}
\hat{\cal H}_{0}(\beta,\lambda) = \frac{1}{W_1} \hat H_{0_{k=1}}
  + \frac{\lambda}{W_k} \hat H_{0_{k}}({\beta}).
\end{equation}
In \Eref{eq:H0}, we have normalized each term with the width of the spectrum 
$W_k$ of the corresponding $k$-body embedded ensemble, which is given 
by~\cite{Asaga2001,BW2003}
\begin{equation}
  \label{width}
  W_k^2 = \frac{1}{N}\overline{{\rm tr}[\hat H_k (\beta)]^2}
           = \Lambda_B^{(0)}(k)+\frac{\delta_{\beta,1}}{N} \sum_{s=0}^k 
               \Lambda_B^{(s)}(n-k),
\end{equation}
where
\begin{equation}
\Lambda_B^{(s)}(k) = {n-s\choose k}{n+s+1\choose k}.
\label{eqLambda}
\end{equation}
These expressions apply to the two single-particle level case ($l=2$); the label 
$B$ stands for bosons, the over-line indicates ensemble average, and 
$\Lambda_B^{(s)}(k)$ is the $s$-th eigenvalue of the ensemble-averaged 
correlation matrix of the bosonic $k$-body embedded ensemble.

We observe that, according to \Eref{eq:H0}, the reference Hamiltonian depends explicitly 
upon $\lambda$. Using the usual creation and annihilation rules, the unperturbed 
energy spectrum can be explicitly calculated
\begin{equation}
\label{eq:E0i}
    E_\mu^{(0)}(\beta,\lambda) =  \frac{\,\epsilon_2 n + (\epsilon_1 - \epsilon_2 )\mu \,}{W_1} 
   + \frac{\lambda}{W_k} \sum_{r=0}^k \, v_{r,r}^{(\beta)} G_{\mu,r}^{(k)} \, .
\end{equation}
Here, $\epsilon_1 > \epsilon_2$ without of loss of generality, and the coefficients 
$G_{\mu,r}^{(k)}$ are identically zero if $r>\mu$ or $k-r>n-\mu$, and otherwise are 
given by
\begin{equation}
  \label{eq:coefs}
    G_{\mu,r}^{(k)} = {\mu \choose r}{n-\mu\choose k-r}\, .
\end{equation}
Then, the residual interaction consists simply of the remaining off-diagonal matrix 
elements of the $k$-body interaction properly normalized 
by $W_k$, i.e., $ \hat{\cal V}^{\beta} =  \hat V_k(\beta)/W_k$.

\subsection{Fidelity and fidelity amplitude}

Fidelity compares the time evolution of a given initial state under a reference 
Hamiltonian, with the time evolution of the same state under a slightly different 
Hamiltonian~\cite{Gorin2006}. We use the Heisenberg time $t_H$ as the time unit, i.e., 
$t = t^\prime/t_H$, where $t_H = 2\pi\hbar/\overline{d}$, and 
$\overline{d}=(\epsilon_1-\epsilon_2)/W_1$ is the mean-level spacing of 
$\hat {\cal H}_0 (\beta,\lambda)$. The unitary time-evolution associated with the 
reference Hamiltonian is given by 
${\cal U}_0^\beta(t)=\hat{T}\exp[- {\rmi 2 \pi / \overline{d}}\, \hat {\cal H}_0 (\beta,\lambda) t]$, 
where ${\hat T}$ is the time-ordering operator. Note that ${\cal U}_0^\beta(t)$ inherits
the $\lambda$ dependence from $\hat{\cal H}_0(\beta,\lambda)$; we drop it from the notation
to make it simpler. We denote by ${\cal U}_\lambda^\beta(t)$ the propagator associated with 
the perturbed Hamiltonian $\hat{\cal H}_\lambda(\beta)$. 

Considering an arbitrary initial state $|\Psi_0\rangle$, the fidelity amplitude is defined as
\begin{equation}
\label{eq:fAmp}
f_{\beta,\lambda}(t) = \langle \Psi_0| M_{\beta,\lambda}(t)  |\Psi_0\rangle = 
\langle \Psi_0| {\cal U}_0^\beta(-t) {\cal U}_\lambda^\beta(t) |\Psi_0\rangle,\ 
\end{equation}
whose square modulus is known as the fidelity
\begin{equation}
\label{eq:fidelity}
F_{\beta,\lambda}(t) = |f_{\beta,\lambda}(t)|^2.
\end{equation}
In~(\ref{eq:fAmp}) we have introduced the echo operator $M_{\beta,\lambda}(t)$, which 
corresponds to the time-evolution propagator associated 
with the time-dependent Hamiltonian $V_{I}(t)$ in the interaction 
picture~\cite{Prosen2002}. 

Following \cite{Prosen2002,Gorin2006}, we use the 
Born expansion of $M_{\beta,\lambda}(t)$, which we truncate at the second 
order; this approximation is referred as the linear response theory. Then, 
\begin{eqnarray}
\label{eq:MAmpO2}
\langle \mu | M_{\beta,\lambda}^{(2)}(t) |\nu\rangle =
   \langle \mu | \Big[ 1 - \rmi w_k \lambda \int_0^t \rmd t_1 V^\beta_{I}(t_1) \nonumber\\
   - w_k^2 \lambda^2
   \int_0^t \rmd t_1 \int_0^{t_1} \rmd t_2\, V^\beta_{I}(t_1) V^\beta_{I}(t_2) \Big] | \nu \rangle ,
\end{eqnarray}
where $V^\beta_{I}(t)= {\cal U}_0^\beta(-t) \hat V_k(\beta) {\cal U}_0^\beta(t)$ also 
depends on $\lambda$ and $\beta$, and $w_k=2 \pi/(\overline{d} W_k)$. We notice 
that this second-order Born expansion contains higher-order contributions with 
respect to $\lambda$, since ${\cal U}^\beta_0(t)$ does depend on $\lambda$. 
Below, we shall restrict to the contributions up to second-order in $\lambda$.

\section{Averaging over the ensemble}
\label{sec:EAverag}

We turn now to the calculation of the ensemble average of the fidelity amplitude. We 
use the occupation-number basis, where the reference Hamiltonian is purely
diagonal and the perturbation has diagonal elements equal to zero. We thus write 
the normalized initial state as $|\Psi_0\rangle = \sum_\mu A_\mu |\mu\rangle$. Then, 
\begin{eqnarray}
  \label{eq:FAmplAv}
  \overline{f_{\beta,\lambda}^{(2)}(t)} = 1 - \rmi w_k \lambda \sum_{\mu,\nu} A_\mu^* A_\nu 
           \int_0^t \rmd t_1 \, \overline{\langle \mu | V^\beta_{I}(t_1) |\nu\rangle} \quad\nonumber\\
           - w_k^2 \lambda^2 \sum_{\mu,\nu} A_\mu^* A_\nu 
           \int_0^t \rmd t_1 \int_0^{t_1} \rmd t_2\,
           \overline{\langle \mu | V^\beta_{I}(t_1) V^\beta_{I}(t_2) | \nu \rangle}\, .\quad
\end{eqnarray}
 
To carry out the calculation, we note that the terms on the r.h.s.~of~(\ref{eq:FAmplAv})
can be factorized in diagonal and off-diagonal contributions of the $k$-body interaction
matrix $v^{(\beta)}$, that is,
\begin{eqnarray}
 \label{eq:Velem1}
 \overline{\langle \mu | V^\beta_{I}(t) | \nu \rangle} & = & \overline{{\cal E}^\lambda_{\mu,\nu}(t)} \,
           \overline{\langle \mu | \hat{V_k}(\beta) | \nu \rangle} \, ,\\
 \label{eq:Velem2}
 \overline{\langle \mu | V^\beta_{I}(t_1) V^\beta_{I}(t_2) | \nu \rangle} & = & \sum_\rho 
      \overline{{\cal E}^\lambda_{\mu,\rho}(t_1) {\cal E}^\lambda_{\rho,\nu}(t_2)}\,\\
 &\times &
   \overline{\langle \mu | \hat{V_k}(\beta) | \rho \rangle \langle \rho | \hat{V_k}(\beta) | \nu \rangle},
   \nonumber
 \end{eqnarray}
where ${\cal E}^\lambda_{\mu,\nu}(t)= \exp\big[\rmi (2 \pi/\overline{d})
(E_\mu^0(\lambda)-E_\nu^0(\lambda))t \big] $. The time-dependence as well as
the additional dependence upon $\lambda$ are contained in the factors 
${\cal E}^\lambda_{\mu,\rho}(t)$, which shows that they are determined by the $k$-body
diagonal elements included in the reference Hamiltonian ${\hat {\cal H}_0}(\beta,\lambda)$.

Using the fact that the $k$-body matrix elements have zero mean implies 
that \Eref{eq:Velem1} is identically zero. For \Eref{eq:Velem2}, we exploit
the fact that the matrix elements of $v^{(\beta)}$ are Gaussian 
distributed variables with fixed variance, i.e., $\overline{v_{r,s}^{(\beta)} v_{t,u}^{(\beta)}}=
(\delta_{r,u} \delta_{s,t} +  \delta_{\beta,1} \delta_{r,t} \delta_{s,u})$. For $\beta=2$
we obtain
\begin{widetext}
  \begin{eqnarray}
  \label{eq:VkVk1}
    \overline{ \langle \mu | \hat V_k ( 2 ) | \rho \rangle \langle \rho | \hat V_k ( 2 ) | \nu \rangle}
      & = & \delta_{\mu,\nu}  \sum_{r\ne s} G_{\mu,r}^{(k)} G_{\rho,s}^{(k)} \, 
    \langle \mu-r | \rho-s \rangle = {\cal G}^{(k)}_{\mu,\rho}\delta_{\mu,\nu} ,\\
    \label{eq:E0iE0i1}
    \overline{ {\cal E}^\lambda_{\mu,\rho}(t_1) {\cal E}^\lambda_{\rho,\nu}(t_2)}
      &=& \exp\Big\{ \rmi 2 \pi \big[(\mu-\rho)t_1+(\rho-\nu)t_2\big] 
        -\frac{1}{2} (w_k \lambda) ^2 \sum_r [ (G_{\mu,r}^{(k)} -G_{\rho,r}^{(k)}) t_1 + 
              (G_{\rho,r}^{(k)}-G_{\nu,r}^{(k)}) t_2 ]^2 \Big\} \, .\qquad
  \end{eqnarray}
\end{widetext}
Here, we denoted by ${\cal G}^{(k)}_{\mu,\rho} = 
\sum_{r\ne s} G_{\mu,r}^{(k)} G_{\rho,s}^{(k)} \langle \mu-r | \rho-s \rangle$, and 
used the fact that $w_{k=1} (\epsilon_1-\epsilon_2) =2\pi$, which follows
from the definitions of $\overline{d}$ and $w_1$ given above. 
For $\mu=\rho$ we have ${\cal G}^{(k)}_{\mu,\mu}=0$, since the matrix 
elements involved are exclusively off-diagonal. For the case $\beta=1$, \Eref{eq:VkVk1} 
has additional contributions which are of order ${\cal O}(1/N)$, and shall be 
neglected.

We note that, due to the first Kr\"onecker delta in \Eref{eq:VkVk1}, the integrand of the second 
term in~(\ref{eq:FAmplAv}) depends only upon the time difference $t_1-t_2$. In addition, 
the second factor of \Eref{eq:E0iE0i1} is a correction due only to the diagonal
$k$-body interactions included in the reference Hamiltonian. Clearly, it is 
Gaussian in $\lambda$, and therefore it contributes to corrections of the order 
of ${\cal O}(\lambda^2)$. Consequently, this factor will induce corrections in
the linear-response formula for the fidelity amplitude at least of the order 
${\cal O}(\lambda^4)$.

\begin{figure}
    \includegraphics[width=8cm]{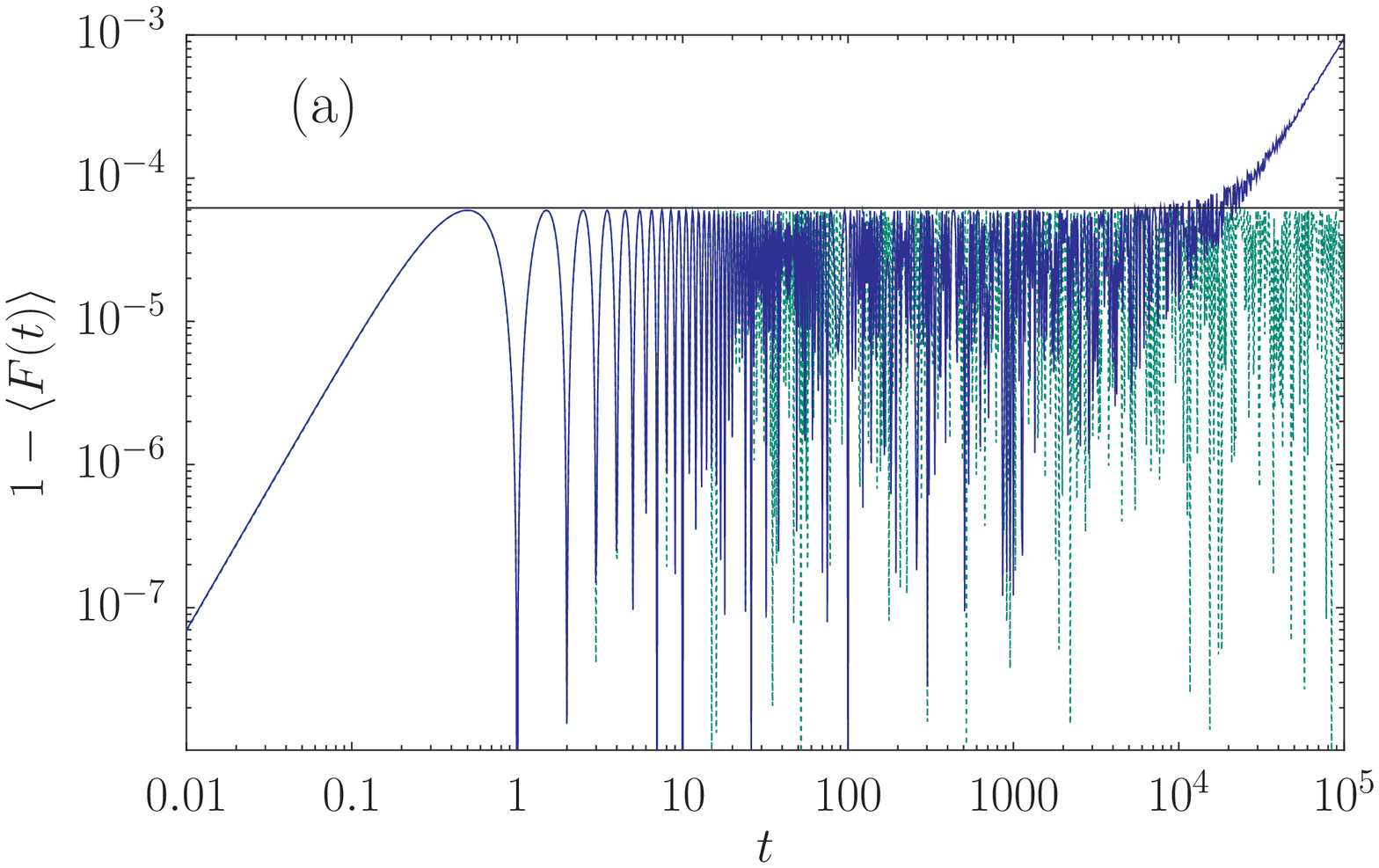}\\
    \includegraphics[width=8cm]{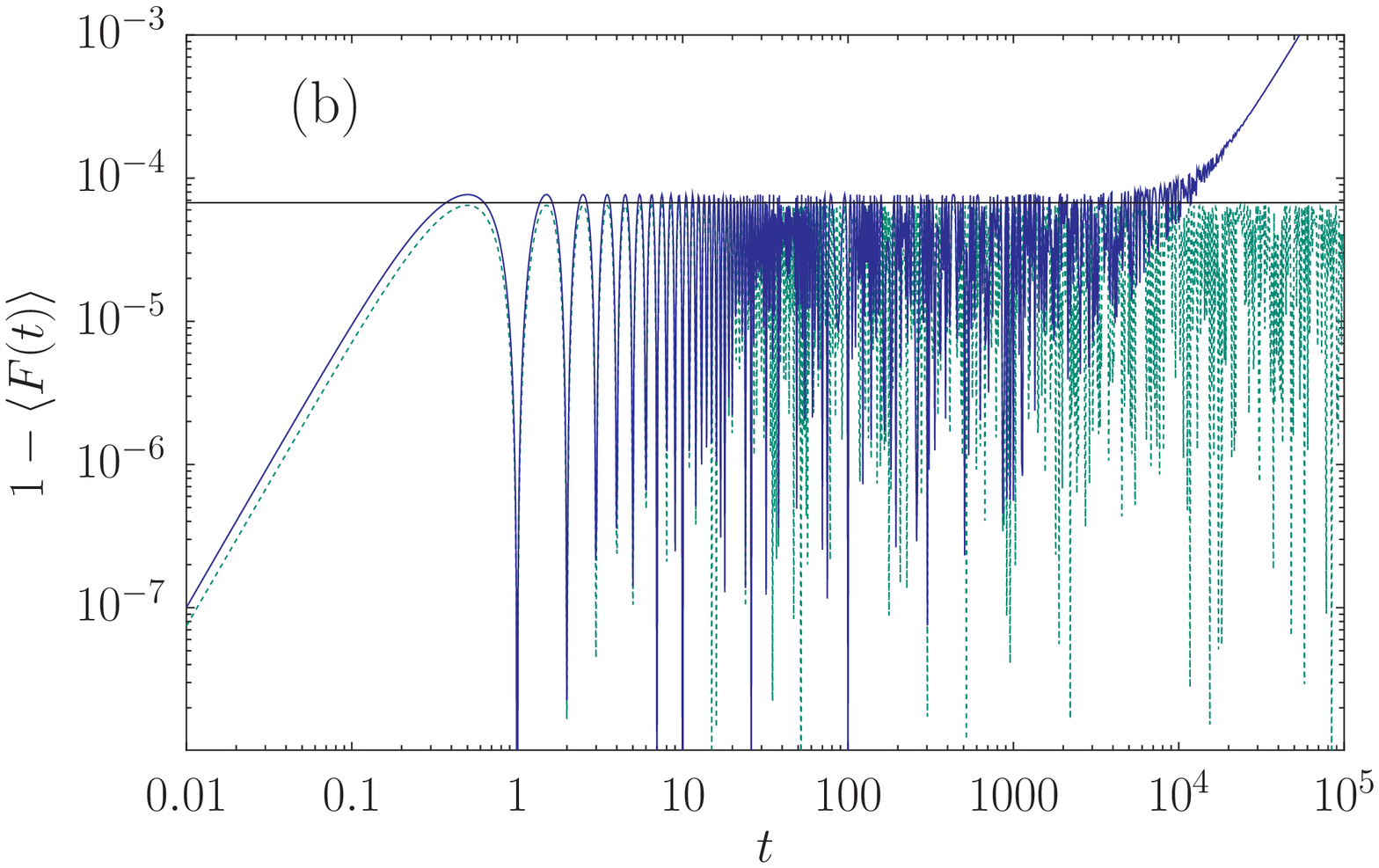}
  \caption{
  Comparison between the ensemble-averaged fidelity given by \Eref{eq:EFid}
  (green/dashed curve) and numerical simulations (continuous/blue curve) 
  for (a)~$\beta=2$ and (b)~$\beta=1$. In this case $k=2$, $\lambda=10^{-6}$, 
  $n=1024$, $\epsilon_1=0.7618036$ and $\epsilon_2=0.9299698$. Notice that the 
  agreement extends to times beyond $10^3 t_H$. The horizontal lines display
  the value of the fidelity freeze according to \Eref{eq:Ffreeze}.
  }
  \label{fig1}
\end{figure}

We consider a normalized random initial state $|\Psi_0\rangle$, and make the
simplification $|A_\mu|^2\sim 1/N$. We carry up the time integrals and compute the
square modulus retaining terms up to second order in $\lambda$. Then, the 
ensemble-averaged fidelity to second order in $\lambda$ is given by
\begin{equation}
\label{eq:EFid}
\overline{F_{\beta,\lambda}^{(2)}(t)} = 1-  \frac{2 w_k^2 \lambda^2}{n+1}  
  \sum_{\mu\neq\rho} {\cal G}^{(k)}_{\mu,\rho}
  \frac{1-\cos[ 2 \pi (\mu-\rho) t  ] }{ [ 2 \pi(\mu-\rho) ]^2}+
  {\cal O}(\lambda^4) .
\end{equation}

\begin{figure}
    \includegraphics[width=8cm]{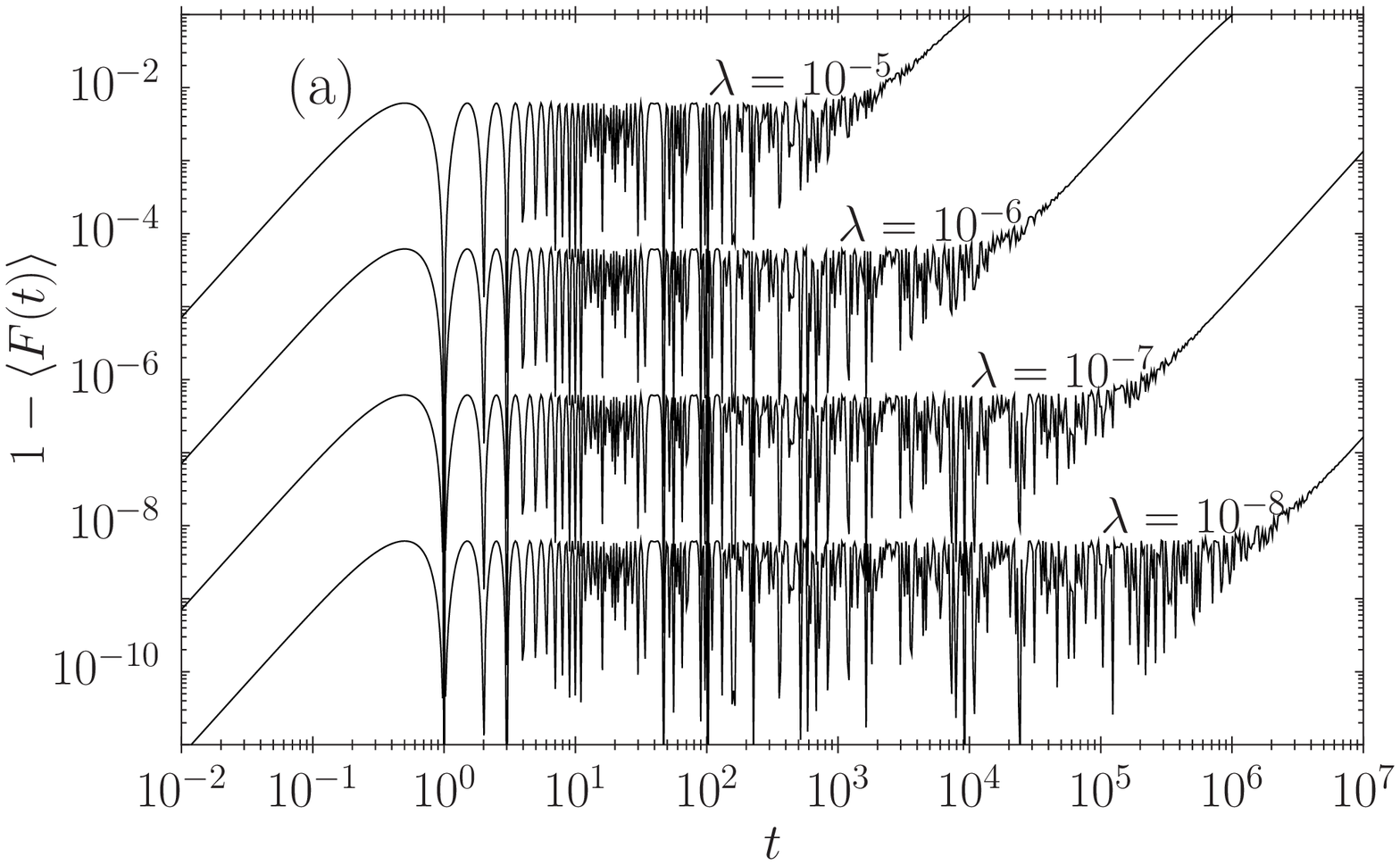}\\
    \includegraphics[width=8cm]{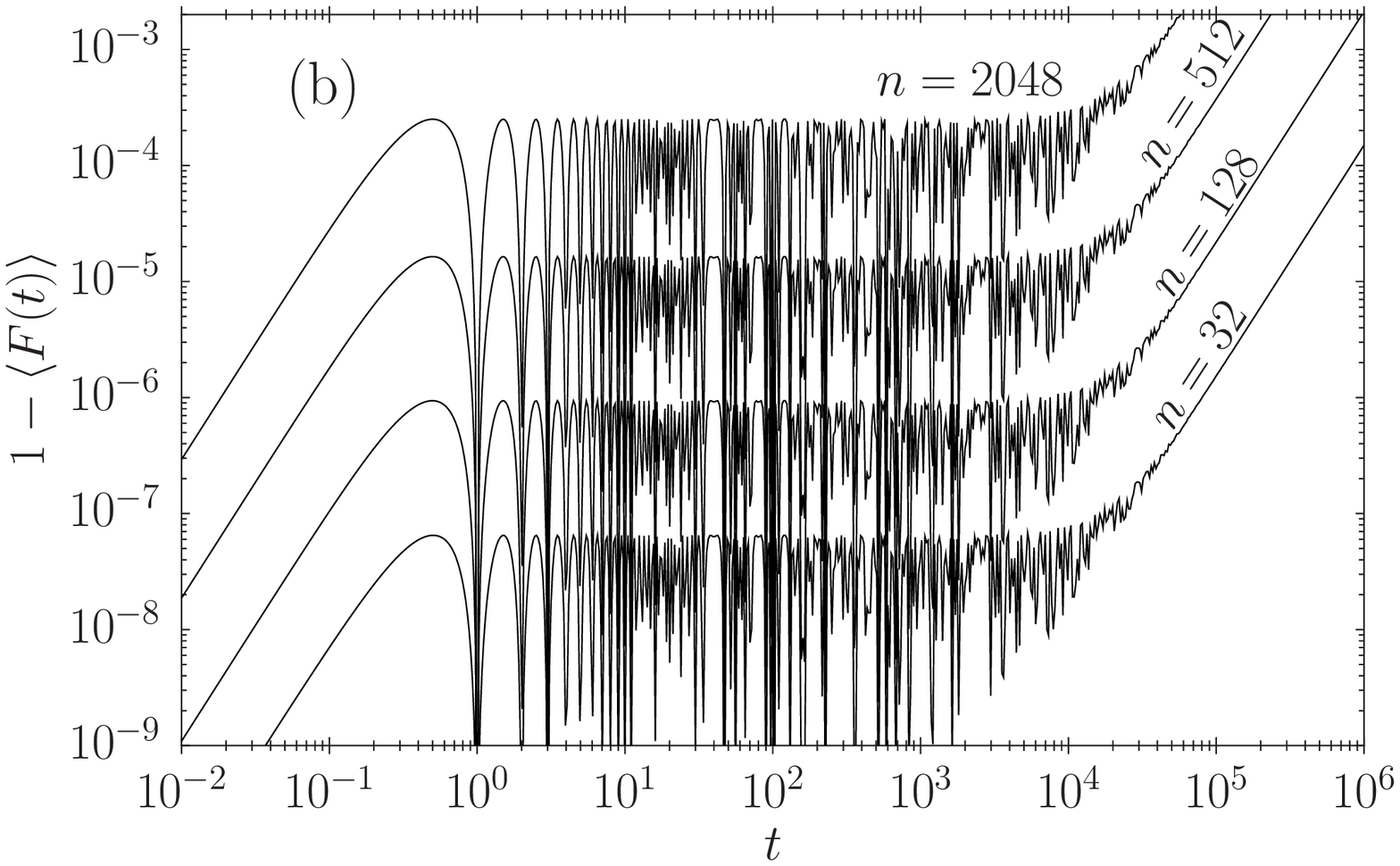}
  \caption{
  Ensemble-averaged fidelity decay for various values of (a)~the perturbation strength 
  $\lambda$, and (b)~the number of particles $n$. Note that the time at which the fidelity 
  freeze ends and decay sets back, $t_e$, is independent of $n$.
  }
  \label{fig2}
\end{figure}

Equation (\ref{eq:EFid}) is the main result in this paper. In \Fref{fig1} we show the comparison 
of the ensemble-averaged fidelity predicted by \Eref{eq:EFid} with numerical 
calculations, both for $\beta=1$ and $\beta=2$. The results show excellent agreement 
even up to rather large values of $t$, when the fourth-order contributions in $\lambda$ 
eventually dominate and destroy the freeze of the fidelity. Notice that the 
agreement also holds for $\beta=1$, in spite of the fact that \Eref{eq:EFid} 
was obtained for $\beta=2$. This confirms {\it a posteriori} that for $\beta=1$ 
the corrections to~(\ref{eq:VkVk1}) are a factor $1/(n+1)$ smaller as assumed, and can 
be neglected in leading order in the boson number. The fact that \Eref{eq:EFid} is 
not a perturbative expansion in time explains the good agreement for large values 
of $t$, which holds as $\lambda$ is small enough. In addition, for small times the 
results confirm the usual quadratic decay of the fidelity, namely,
$1-\overline{F_{\beta,\lambda}^{(2)}(t)} \propto t^2 + {\cal O}(t^4)$. 
\Fref{fig2} displays the dependence of the ensemble-averaged fidelity 
with respect to $\lambda$ or $n$.

More interesting and far reaching is the fact that fidelity, up to second order in 
$\lambda$ is a Fourier cosine-series in $t$. The basic periodicity is precisely the 
Heisenberg time $t_H=1$, since the minimum difference in the 
occupation numbers for the states $|\mu\rangle$ and $|\rho\rangle$ in one of the 
single-particle states is precisely 1. Indeed, the double sum over the basis states 
excludes the case $\mu=\rho$, since for $\mu=\rho$ we have set 
${\cal G}^{(k)}_{\mu,\mu}=0$ before the time-integration is carried up.
This emphasizes the fact that the Fourier coefficients in \Eref{eq:EFid} are related to the 
off-diagonal residual interaction. Moreover, we observe that, in units of the $t_H$, 
at integer values of time the time-dependent term of~(\ref{eq:EFid}) vanishes identically, 
and therefore revival of $|\Psi_0\rangle$ are observed, as illustrated in~ \Fref{fig1}. These 
revivals are not a full recovery of $|\Psi_0\rangle$ though, since there 
may be corrections of higher order in $\lambda$, and \Eref{eq:EFid} has been 
averaged over the ensemble. We emphasize that the periodicity $T_f=t_H$ of 
the revivals of the ensemble-averaged fidelity for the bosonic embedded ensembles
follows from the fact that $\epsilon_1\neq\epsilon_2$, which is responsible 
for the complex exponential factor of \Eref{eq:E0iE0i1}. 

\begin{figure}
    \includegraphics[width=8.8cm]{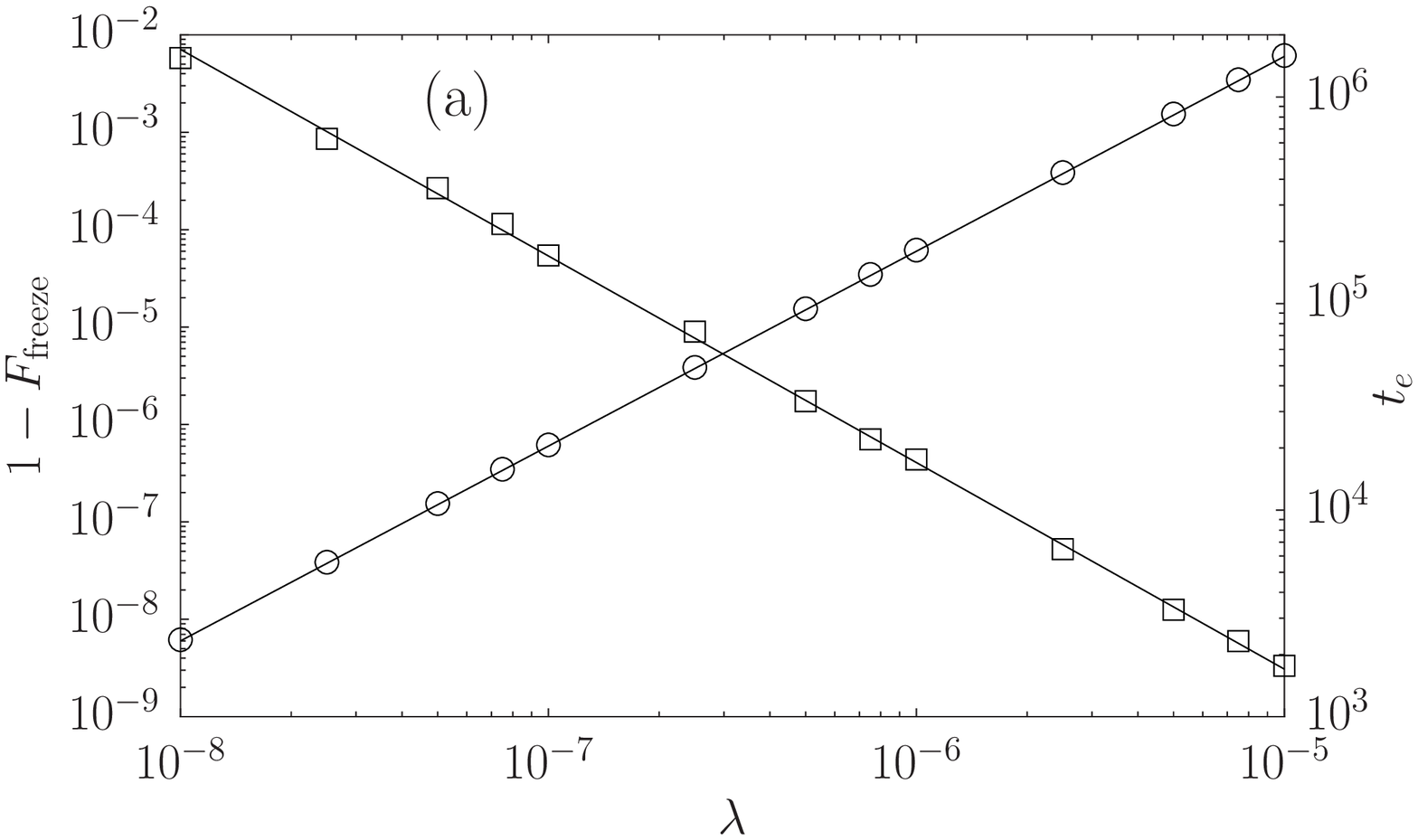}\\
    \includegraphics[width=8.0cm]{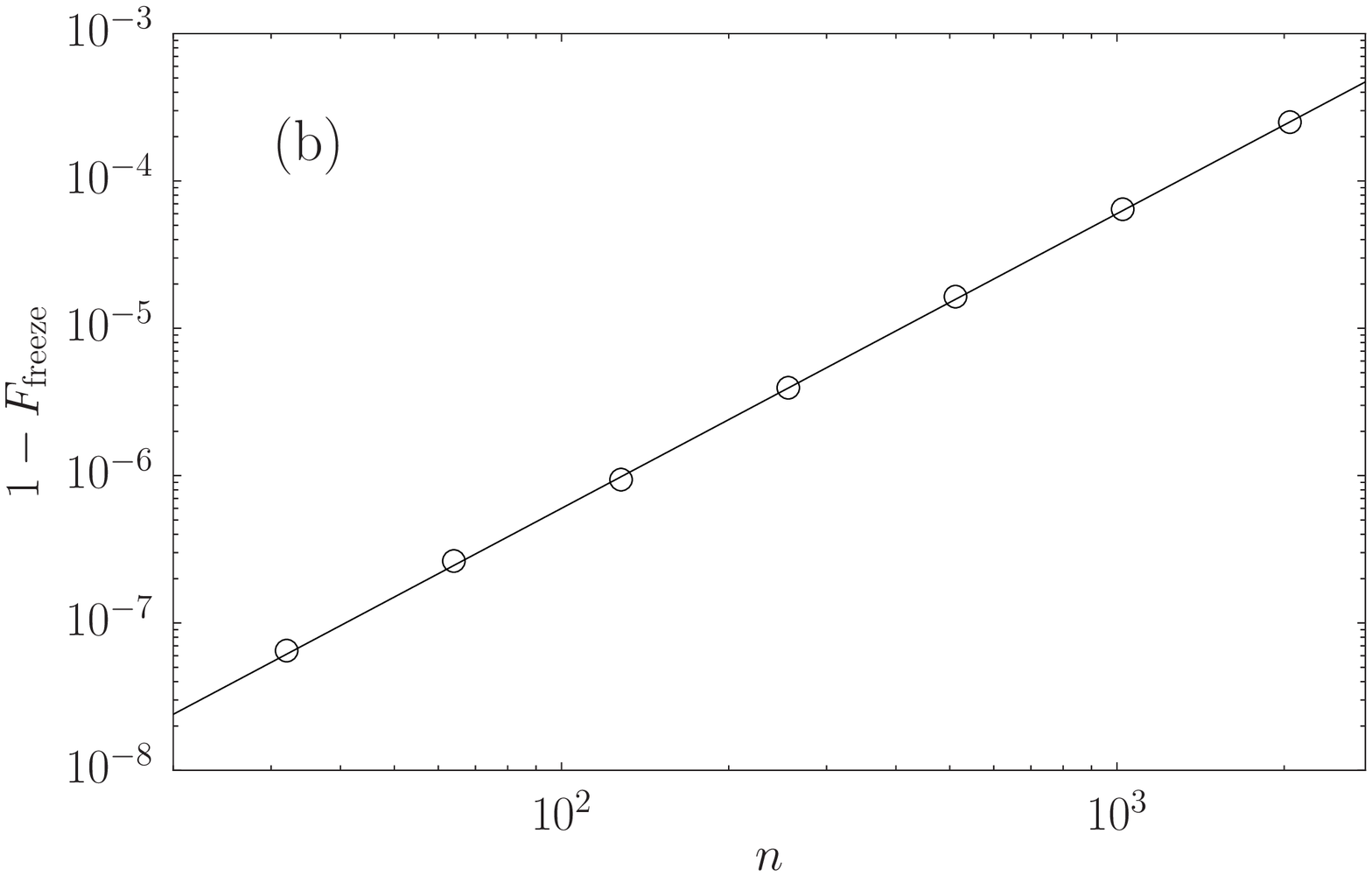}
  \caption{
  (a) Dependence of the freeze of the fidelity (circles) upon $\lambda$, which
  confirms that $1-F_{\rm freeze} \propto \lambda^2$. The right vertical
  scale indicates the time of duration of the freeze of the fidelity (data in
  squares) which shows the scaling $t_e\propto 1/n$. (b)~Dependence of 
  $F_{\rm freeze}$ on $n$ confirming the scaling $1-F_{\rm freeze} \propto n^2$. }
  \label{fig3}
\end{figure}

Equation~(\ref{eq:EFid}) permits to obtain an estimate of the freeze of the fidelity, which 
we denote 
by $F_{\rm freeze}$. Considering the minimum value for the fidelity when~(\ref{eq:EFid})
is valid, that is, when the periodic revivals are observed because of the freeze of the 
fidelity, we have
\begin{equation}
\label{eq:Ffreeze}
F_{\rm freeze} = 1 - \frac{(w_k \lambda)^2}{\pi^2 (n+1)} \sum_{\mu\neq\rho}
  \frac{  {\cal G}^{(k)}_{\mu,\rho}}{ (\mu-\rho)^2 }.
\end{equation}
Equation~(\ref{eq:Ffreeze}) predicts that $F_{\rm freeze}$ scales as $\lambda^2$. The scaling 
with respect to the number of particles is more involved since $w_k$ and
the coefficients ${\cal G}_{\mu,\rho}^{(k)}$ depend on $n$. Using Stirling's 
formula we obtain $w_k^2\sim n^{2-2k}$ and ${\cal G}_{\mu,\rho}^{(k)} \sim n^{2k}$,
which yield the scaling $\sim n^2$, where we took into account that the sum over the 
many-body states cancels the normalization factor of the random initial state 
$|\Psi_0\rangle$. These scaling laws are confirmed numerically as illustrated in 
 \Fref{fig3}. At this point we note that the time $t_e$ during which the freeze of the fidelity 
lasts scales as $t_e\sim\lambda^2$ and is essentially independent of $n$; cf. 
Figs.~\ref{fig2} and~\ref{fig3}.

\section{Periodic fractional revivals and $k$-body interactions}
\label{sec:PFrevivals}

As discussed above, \Eref{eq:EFid} predicts periodic time-revivals of fixed period 
$T_f=t_H=1$, independently of the rank $k$ of the residual interaction. This is a 
consequence of the fact that the residual $k$-body interaction so far considered, 
contains terms which involve moving $1, 2, \dots, k$  particles, from one of the 
single-particle levels to the other. That is, the many-body states coupled through 
the perturbation $\hat V_k(\beta)$ may differ at least in the occupation of one
particle, and at most in the occupation of $k$. These differences are precisely the 
factors $\mu-\rho$ that appear in the Fourier expansion in \Eref{eq:EFid}, which 
denote the difference of number of particles in a given single-particle level. Clearly,
the minimum difference fixes the periodicity of the Fourier cosine series.

\begin{figure}
  \includegraphics[width=8.2cm]{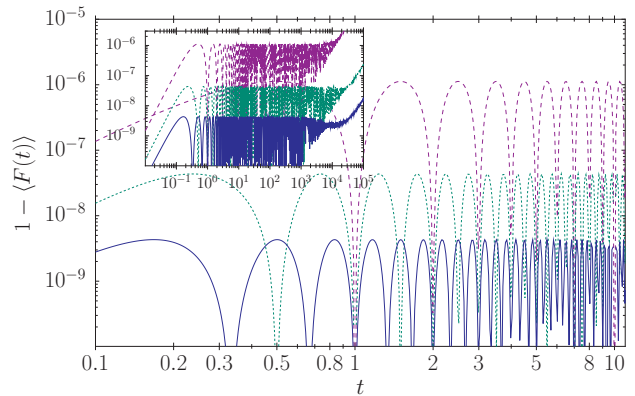}
  \caption{
  Ensemble-averaged fidelity decay for the residual interaction $\hat{\cal K}_k$
  in \Eref{eq:HkMod} for $k=2$ (green/dotted curve) and $k=3$ (blue/continuous curve), 
  illustrating the fractional periodic revivals. These correspond to $T_f=1/2$ and 
  $T_f=1/3$, respectively. The magenta/dashed curve displays the result when all 
  off-diagonal elements for $k=3$ of (\ref{eq:Hk}) are considered.  }
  \label{fig4}
\end{figure}

This explanation opens the following interesting possibility. By selecting 
the actual perturbing terms within the $k$-body residual interaction, we can actually 
tune the observed periodicity of the revivals during the fidelity freeze, in particular, 
making it different from Heisenberg time $T_f=1$. Indeed, we can fix the perturbation 
such that the only terms present move exactly $k$ particles from one single-particle 
level to the other one. That is, we restrict the general $k$-body Hamiltonian 
(\ref{eq:Hk}) such that the off-diagonal terms are
\begin{equation}
  \label{eq:HkMod}
   \hat{\cal K}_k =  v_{k,0} (\hat a_1^\dagger)^k (\hat a_2)^k + 
   v_{0,k} (\hat a_2^\dagger)^k (\hat a_1)^k ,
\end{equation}
with $v_{k,0}=(v_{0,k})^*$ to ensure hermitecity. This choice of the residual interaction 
implies that the Fourier coefficients vanish unless $|\mu-\rho|=k$, that is, when the 
states $|\mu\rangle$ and $|\rho\rangle$ precisely differ in the occupation of $k$
particles with respect to one mode. In this case, the argument of the cosine 
function in \Eref{eq:EFid} is $2\pi k t$, which implies that the periodicity of the 
revivals during the freeze of the fidelity becomes $T_f=1/k$ in units of the 
Heisenberg time. Apart from the fact that this periodicity differs from $1$, the 
important aspect is that the periodicity of the revivals during the freeze of the 
fidelity provides a direct measure of the rank of the interaction $k$ of the 
residual perturbation $\hat{\cal K}_k$. These are the fractional periodic revivals.

This prediction is illustrated in  \Fref{fig4}, where we plot
the ensemble-averaged fidelity for an interaction of the form (\ref{eq:HkMod})
for $k=2$ and $k=3$, and for comparison a case including all off-diagonal
contributions of $\hat H_k$ in \Eref{eq:Hk}. The results clearly show $T_f=1/2$, $T_f=1/3$ 
for the first two cases, reflecting the value of the corresponding $k$-body interactions,
and $T_f=1$, as expected from \Eref{eq:EFid}. 

\section{Conclusions}
\label{sec:Concl}

In this paper, we have studied analytically and numerically the fidelity decay in the 
$k$-body embedded ensemble of random matrices for bosons distributed in two 
single-particle levels. We defined fidelity in terms of a reference Hamiltonian, which 
is assumed to be in diagonal form, and a perturbed Hamiltonian which in addition 
includes a purely off-diagonal residual $k$-body interaction. This situation mimics 
the typical set-up in mean-field calculations, but appears also in other interesting 
physical cases, such as time-reversal symmetry-breaking. This set-up fulfills 
the conditions to observe the freeze of the fidelity~\cite{ProsenZnid2003,GKPSSZ2006}, 
thus allowing for longer control of the system.

We calculated the ensemble-averaged fidelity within the linear response theory up to 
second-order in the perturbation parameter, which is a Fourier series whose basic 
periodicity is equal to the Heisenberg time $t_H$. The analytical predictions are in good 
correspondence with the direct numerical results, confirming the presence of the freeze 
for the ensemble-averaged fidelity as well as the relevant scalings with respect to the 
strength of the perturbation and the number of particles of the system. The oscillatory
part of this Fourier series cancels at integer times of $t_H$, thus manifesting the
periodicity of the revivals. Selecting the off-diagonal terms of the $k$-body residual 
interaction, in order that the actual perturbation couples only many-body states differing 
exactly by $k$ particles in the occupation number of either single-particle level, we 
showed that the periodicity of the revivals becomes $1/k$ in units of the Heisenberg 
time. Therefore, the periodicity of the revivals of the ensemble-average fidelity during 
freeze may be used as a direct measure to detect the rank $k$ of the perturbing 
interaction. This aspect may be interesting in the context of current efforts that address 
effects related to three--body interactions~\cite{Buechler2007,Johnson2009} which 
are responsible, for instance, for atomic losses in ultra-cold bosonic gases.

\begin{acknowledgments}
We acknowledge financial support from the projects IN-114310 (DGAPA-UNAM) 
and 57334-F (CONACyT). LB is thankful to D. Sahag\'un for discussions and 
correspondence, and the kind hospitality of \`A. Jorba and C. Sim\'o at the U. of
Barcelona, where this work was completed. LB acknowledges financial support 
from Programa de Estancias Posdoctorales y Sab\'aticas (CONACyT)
\end{acknowledgments}



\begin{thebibliography}{99}

\bibitem{Gorin2006}%
  T. Gorin, T. Prosen, T.H. Seligman, and M. \v{Z}nidari\v{c}, {Phys. Rep} {\bf 435}, 33 (2006).

\bibitem{ProsenZnid2003}
  T. Prosen and M. \v{Z}nidari\v{c}, {New J. Phys. } {\bf 5}, 109 (2003);
  T. Prosen and M. \v{Z}nidari\v{c}, {Phys. Rev. Lett.} {\bf 94}, 044101 (2005).

\bibitem{Pizorn2007}
  I. Pi\v{z}orn, T. Prosen and T.H. Seligman, {Phys. Rev. B} {\bf 76} 035122 (2007).

\bibitem{Lobkis2003}
  O.I. Lobkis and R.L. Weaver, {Phys. Rev. Lett.} {\bf 90}, 254302 (2003).
  
\bibitem{GSW2006}
  T. Gorin, T.H. Seligman and R.L. Weaver, {Phys. Rev. E} {\bf 73}, 015202(R) (2006).

\bibitem{BJL2003}
  L. Benet, C. Jung and F. Leyvraz, {J. Phys. A: Math. Gen.} {\bf 36}, L217 (2003).

\bibitem{Anderson1998}
  B.P. Anderson, M.A. Kasevich, {Science} {\bf 282}, 1686 (1998).

\bibitem{Albiez2005}%
  M. Albiez, R. Gati, J. F\"olling, S. Hunsmann, M. Cristiani, and
  M. K. Oberthaler, {Phys. Rev. Lett.} {\bf 95}, 010402 (2005).

\bibitem{Morsch2006}%
  O. Morsch and M. Oberthaler, {Rev. Mod. Phys.} {\bf 78}, 179 (2006).

\bibitem{Gati2007}%
  R. Gati, M.K. Oberthaler, {J. Phys. B: At. Mol. Opt. Phys.} {\bf 40}, R61 (2007).

\bibitem{fre70}
  J.B. French and S.S.M. Wong, {Phys. Lett. B} {\bf 33}, 449 (1970);
  J.B. French and S.S.M. Wong, {Phys. Lett. B} {\bf 35}, 5 (1971).

\bibitem{boh71}
  O. Bohigas and J. Flores, {Phys. Lett. B} {\bf 34}, 261 (1971),
  O. Bohigas and J. Flores, {Phys. Lett. B} {\bf 35} 383 (1971).

\bibitem{mon75}
  K.K. Mon and J.B. French, {Ann. Phys. (N.Y.)} {\bf 95}, 90 (1975).

\bibitem{kot80} 
  V.K.B. Kota and V. Potbhare, {Phys. Rev. C} {\bf 21}, 2637 (1980).

\bibitem{man84}
 V.R. Manfredi, {Lett. Nuovo Cimento} {\bf 40}, 135 (1984).

\bibitem{Asaga2001}%
  T. Asaga, L. Benet, T. Rupp and H.A. Weidenm\"uller, {Eurphys. Lett.} {\bf 56}, 340 (2001);
  {\it ibid}, {Ann. Phys.} (N.Y.) {\bf 298}, 229 (2002).

\bibitem{Pizorn2008}
  I. Pi\v{z}orn, T. Prosen, S. Mossmann and T.H Seligman, {New J. Phys.} {\bf 10}, 
  023020 (2008).

\bibitem{Kota2001}
  V.K.B. Kota, {Phys. Rep.} {\bf 347}, 223 (2001).
  
\bibitem{BW2003}
  L. Benet and H.A. Weidenm\"uller, {J. Phys. A: Math. Gen.} {\bf 36}, 3569 (2003).

\bibitem{BHQS2010}
  L. Benet, S. Hern\'andez-Quiroz and T.H. Seligman, {AIP Conf. Proc.}
  {\bf 1323}, 6 (2010).

\bibitem{BLS2003}%
  L. Benet, F. Leyvraz and T.H. Seligman, {Phys. Rev. E} {\bf 68},
  045201(R) (2003).

\bibitem{HQ-B2010}%
  S. Hern\'andez-Quiroz and L. Benet, {Phys. Rev. E} {\bf 81}, 036218 (2010).

\bibitem{GMGW1998}%
  T. Guhr, A. Mueller-Gr\"oling and H.~A. Weidenm\"uller, {Phys. Rep.} 
  {\bf 299}, 189 (1998).

\bibitem{Shnirelman1975}%
  A. I. Shnirelman, Usp. Mat. Nauk {\bf 30}, 265 (1975);
  A. I. Shnirelman, addendum in V. F. Lazutkin, KAM Theory and
  Semiclassical Approximations to Eigenfunctions (Springer, Berlin,
  1993).

\bibitem{GKPSSZ2006}%
  T. Gorin, {\it et al.}, 
  {Phys. Rev. Lett.} {\bf 96}, 244105 (2006).

\bibitem{Prosen2002}
  T. Prosen, {Phys. Rev. E} {\bf 65}, 036208 (2002);
  T. Prosen and M. \v{Z}nidari\v{c}, {J. Phys A: Math Gen} {\bf 35}, 1455 (2002).

\bibitem{Buechler2007}%
  H.P. B\"uchler, A. Micheli and P. Zoller, Nature Physics {\bf 3} 726 (2007).

\bibitem{Johnson2009}%
  P. R. Johnson, E. Tiesinga, J. V. Porto and C. J. Williams, New
  Journal of Physics {\bf 11}, 093022 (2009), and references therein.


\end{thebibliography}
\end{document}